\documentclass[aps, pra, amsmath, showpacs, preprintnumbers, superscriptaddress, twocolumn, amssymb]{revtex4-1}
\usepackage{CJK}
\usepackage{braket}
\usepackage{graphicx}
\usepackage{mathptmx, textcomp}
\usepackage[latin1]{inputenc}
\usepackage{tabularx}
\usepackage[usenames]{color}
\usepackage[colorlinks, urlcolor=blue, citecolor=blue, linkcolor=blue]{hyperref}
\begin{document}
\begin{CJK*}{UTF8}{gbsn} % Use default fonts from CJK for Chinese (see below)
\title{Finite-temperature effects on a triatomic Efimov resonance in ultracold cesium}
\author{B. Huang (黄博)} 
\author{L. A. Sidorenkov}
\altaffiliation{Present address: Laboratoire Kastler Brossel, Coll\`ege de France, CNRS, ENS-PSL Research
University, UPMC-Sorbonne Universit\'es, 75005 Paris.}
\author{R. Grimm}
%\affiliation{Institut f\"ur Experimentalphysik Universit\"at
% Innsbruck, % Technikerstra{\ss}e 25,
% 6020 Innsbruck, Austria\\
%and Institut f\"ur Quantenoptik und Quanteninformation (IQOQI),
% \"Osterreichische Akademie der Wissenschaften, 6020 Innsbruck,
% Austria}
\affiliation{Institut f\"ur Experimentalphysik, Universit\"at
 Innsbruck, % Technikerstra{\ss}e 25,
 6020 Innsbruck, Austria}
\affiliation{Institut f\"ur Quantenoptik und Quanteninformation (IQOQI),
 \"Osterreichische Akademie der Wissenschaften, 6020 Innsbruck,
 Austria}

\date{\today}

\begin{abstract}
We report a thorough investigation of finite-temperature effects on three-body recombination near a triatomic Efimov resonance in an ultracold gas of cesium atoms. Our measurements cover a wide range from a near-ideal realization of the zero-temperature limit to a strongly temperature-dominated regime. The experimental results are analyzed within a recently introduced theoretical model based on a universal zero-range theory. The temperature-induced shift of the resonance reveals a contribution that points to an energy-dependence of the three-body parameter. We interpret this contribution in terms of the finite range of the van der Waals interaction in real atomic systems and we quantify it in an empirical way based on length scale arguments. A universal character of the corresponding resonance shift is suggested by observations related to other Efimov resonances and the comparison with a theoretical finite-temperature approach that explicitly takes the van der Waals interaction into account. Our findings are of importance for the precise determination of Efimov resonance positions from experiments at finite temperatures.
\end{abstract}

\maketitle
\end{CJK*}

%% introduction
\section{Introduction}
Few-body quantum physics with ultracold atoms has emerged as a new research field, connecting basic concepts from nuclear, molecular, and atomic physics \cite{Braaten2006uif, Ferlaino2010fyo, Wang2013ufb}. The paradigm of the field is Efimov's prediction of universal three-body states \cite{Efimov1970ela}. Efimov  showed that, when two bosons interact with an infinite scattering length, the corresponding three-particle system has an infinite number of three-body states just below
threshold.  Signatures of Efimov states were first observed in an ultracold gas of cesium atoms
\cite{Kraemer2006efe}, and have since been found in many other ultracold systems,
including other bosonic gases \cite{Zaccanti2009ooa, Pollack2009uit,
Gross2009oou, Gross2010nsi, Wild2012mot, Roy2013tot}, three-component fermionic
spin mixtures \cite{Ottenstein2008cso, Huckans2009tbr, Williams2009efa,
Nakajima2010nea}, and mixtures of atomic species \cite{Barontini2009ooh,
Bloom2013tou, Pires2014ooe, Tung2014gso}. Very recently, the existence of an Efimov state has also been confirmed for helium atoms in a molecular beam \cite{Kunitski2015oot}. Moreover, extensions of the Efimov scenario to universal states of larger clusters \cite{Hammer2007upo, vonStecher2009sou, vonStecher2010wbc} have been demonstrated in
experiments \cite{Ferlaino2009efu, Pollack2009uit, Zenesini2013rfb},
highlighting the general nature of universal few-body physics.

The universal regime of few-body physics is realized when the $s$-wave scattering length $a$ is well separated from all other length scales of the problem.  This means that $a$ has to be large compared with the relevant range of the two-body  interaction potential, but small compared with the thermal de Broglie wavelength of the sample.  Thus, the conceptually most simple case is the idealized scenario of a zero-range two-body interaction in a zero-temperature ensemble. This case, which has been widely discussed in the literature, is commonly referred to as the ``universal limit'' of few-body physics, where all observables are uniquely connected by fixed relations \cite{Braaten2006uif, Wang2013ufb}. Many experiments have focused on tests of these universal relations, concerning the famous Efimov period \cite{Zaccanti2009ooa, Pollack2009uit, Dyke2013frc, Huang2014oot}, the relation between features at positive and negative scattering lengths \cite{Kraemer2006efe, Knoop2009ooa, Zaccanti2009ooa, Zenesini2014rad}, and the relation between three-body and $N$-body resonances \cite{Ferlaino2009efu,  Pollack2009uit, Zenesini2013rfb}. Some of the experiments \cite{Knoop2009ooa, Dyke2013frc, Huang2014oot, Zenesini2014rad} have revealed deviations from perfect universality, which challenge our understanding of the intricate connections between the idealized few-body scenario and systems that exist in the real world. 

Efimov resonances in three-body loss \cite{Ferlaino2011eri} represent the main observables in few-body physics with ultracold atoms. They mark the points where the three-atom states cross the dissociation threshold. The finite temperature of the ensemble shifts those resonances as the Efimov state then couples to the scattering continuum, and the feature turns into a triatomic continuum resonance \cite{Bringas2004tcr}. The limitations by unitarity \cite{Dincao2004lou, Rem2013lot, Fletcher2013soa} lead to a saturation of the maximum resonance amplitude, and the resulting effect is a blurring with a loss of visibility \cite{Jonsell2006esf, Massignan2008esn, Wang2014uvd}. Nevertheless, even when an experimental observation is strongly influenced by the finite temperature, an appropriate theoretical model allows one to extract the zero-temperature position of an Efimov resonance. Knowledge of this position also determines the three-body parameter (3BP) \cite{Braaten2006uif, Wang2013ufb}, which fixes the ladder of Efimov states.

In our recent work on cesium \cite{Huang2014oot},  we have observed and analyzed an {\em excited-state} Efimov resonance. Such higher-order resonances are, under realistic experimental conditions, always strongly influenced by the temperature. We have employed the theoretical model of Ref.~\cite{Rem2013lot} to extract the resonance position for the zero-temperature limit. The theoretical approach is based on an $S$-matrix formalism and provides a non-perturbative solution for any temperature under the basic assumption of a two-body interaction with a zero range. We thus refer to it as the universal zero-range (UZR) model. An extension of the theory has been applied in Ref.~\cite{Huang2014tbp} to analyze observations of an excited-state Efimov resonance in the three-fermion system of $^6$Li \cite{Williams2009efa}. A further extension has been presented in Ref.~\cite{Petrov2015tbr} for mass-imbalanced three-body systems, which are in the focus of current experimental work  \cite{Pires2014ooe, Tung2014gso, Ulmanis2015uow}.

In this Article, we investigate a {\em ground-state} Efimov resonance in cesium over a wide temperature range. Our measurements cover conditions from a near-ideal realization of the zero-temperature limit to a strongly temperature-dominated regime. The experimental results are analyzed within the framework of the UZR theory. This reveals a temperature-dependent resonance shift that corresponds to a variation of the 3BP with the collision energy. For a real atomic system, we introduce the length that is associated with the van der Waals attraction and characterize the shift in an empirical way based on length scale arguments. This improves the accuracy of the determination of the zero-temperature resonance position (and thus of the 3BP in the zero-energy limit) from experimental data. In Sec.~\ref{sect:role-of-T}, we discuss different finite-temperature regimes relevant for the experiments in Cs. In Sec.~\ref{sect:experiments}, we present our experimental results together with an analysis based on the UZR model. In Sec.~\ref{sect:discussion}, we discuss our findings in view of previous and future experiments in the field.

\section{Finite-temperature regimes}\label{sect:role-of-T}

\begin{figure}
\includegraphics[width=8.5cm]{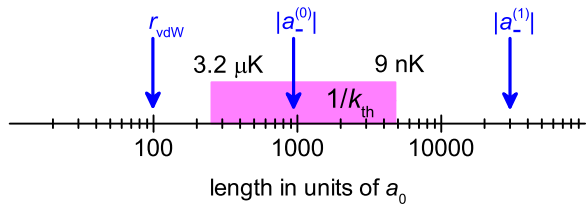}
\vspace{-3mm}
\caption{(Color online) Length scales involved in experiments on Efimov states in ultracold cesium gases. The parameters $a_-^{(0)}$ and $a_-^{(1)}$ denote the zero-temperature positions of the resonances associated with the Efimov ground state and the first excited state, respectively. The shortest length scale is the van der Waals length $r_{\rm vdW}$. The temperature is described by the corresponding length $1/k_{\rm th}$ (see text). The shaded region indicates the temperature range explored in our Cs experiments, which extends from 9~nK to 3.2~$\mu$K, corresponding to $1/k_{\rm th} $ from $4800\,a_0$ to $250\,a_0$.} \label{fig:LengthScales}

\end{figure}

Different regimes in experiments on Efimov states can be discussed in terms of length scales. Figure~\ref{fig:LengthScales} illustrates the situation for the Cs atom. The shortest relevant length scale is the van der Waals length $r_{\rm vdW} = \frac{1}{2} (m C_6 / \hbar^2)^{1/4}$ \cite{Chin2010fri}, where $m$ is the atomic mass and $C_6$ is the van der Waals coefficient. This length quantifies the long-range part of the two-body interatomic potential, and for Cs $r_{\rm vdW} = 101.1\,a_0$ with $a_0$ denoting Bohr's radius. The Efimov ground-state is characterized by the parameter $a_-^{(0)}$, which specifies the scattering length at which an Efimov state crosses the dissociation threshold. %corresponding to the point where the ground-state Efimov resonance appears in the zero-temperature limit. 
This parameter, which corresponds to the 3BP in the zero-energy limit, also characterizes the typical size of the Efimov ground state. Recent work has revealed a new kind of `3BP universality' \cite{Berninger2011uot, Wang2012oot, Schmidt2012epb, Sorensen2012epa, Naidon2014moa,
Wang2014uvd}, which is specific to atomic systems and links this parameter to the van der Waals length by $a_-^{(0)}  \approx -9.5\, r_{\rm vdW}$. The first excited Efimov state is characterized by the analogously defined resonance position $a_-^{(1)} \approx 22.7 a_-^{(0)}$. 

To tune the scattering length in Cs, a broad Feshbach resonance near 800\,G serves as an excellent tool \cite{Berninger2011uot, Berninger2013frw}. This resonance represents the most extreme case of an entrance-channel dominated resonance \cite{Chin2010fri} that is known for any species. Here the scattering problem can be described in terms of an effective single-channel model, neglecting the intrinsic two-channel nature of a Feshbach resonance. For this resonance, the length parameters $a_-^{(0)} =  -963(11)\,a_0$ and  $a_-^{(1)} =  - 20190(1200)\,a_0$ have been determined \cite{Huang2014oot}, corresponding to an Efimov period of $a_-^{(1)}/a_-^{(0)} = 21.0(1.3)$. The first observation of an Efimov resonance in Cs \cite{Kraemer2006efe} was made at low magnetic fields near 7.5\,G, where a Feshbach resonance of similar character exists. Here $a_-^{(0)} =  -872(22)\,a_0$ was found, which is very close to the observation on the high-field resonance. 

We now introduce the thermal length scale $1/k_{\rm th}$, where $k_{\rm th}=\sqrt{2\pi mk_BT}/\hbar$ is the thermal de Broglie wavenumber. Experiments on Cs have been carried in a temperature range between $9\,$nK \cite{Huang2014oot} and 3.2\,$\mu$K. For the lowest temperature, $T = 9\,$nK, the length $1/k_{\rm th} = 4800\,a_0$ is right between $|a_-^{(0)}|$ and $|a_-^{(1)}|$. This means that the ground-state Efimov resonance will show very little temperature effects since the dimensionless parameter $k_{\rm th} |a_-^{(0)}| = 0.26$ is quite small, which indeed was the case in the experiments  reported in Refs.~\cite{Kraemer2006efe, Berninger2011uot}. In contrast, the excited-state Efimov resonance will be strongly dominated by finite-temperature effects since $k_{\rm th} |a_-^{(1)}| \approx 4$ is rather large, which was the case in the experiments of Ref.~\cite{Huang2014oot}. In an intermediate experimental temperature regime, with $T \approx 220$\,nK  one obtains $k_{\rm th} |a_-^{(0)}| \approx 1$, which means that also the ground-state Efimov resonance will be subject to substantial temperature effects. This intermediate  regime was already investigated in Ref.~\cite{Kraemer2006efe}.  

At the highest experimental temperatures of $3.2\,\mu$K (this work) we realize $k_{\rm th} |a_-^{(0)}| \approx 4$, so that strongly temperature-dominated behavior of the ground-state resonance can be expected, quite similar to the excited-state Efimov resonance at 9\,nK. However, in this regime, an additional complication arises as even the smallest dimensionless parameter in the problem, $k_{\rm th} r_{\rm vdW}$, is no longer small. This suggests the appearance of a combined effect of finite temperature and finite range.

\section{Experiments}\label{sect:experiments}
In this Section, we present our experimental results for the temperature dependence of the ground-state Efimov resonance that appears in Cs near a magnetic field of 853~G~\cite{Berninger2011uot}. For an accurate conversion of the magnetic field to the scattering length we employ the model of Ref.~\cite{Berninger2013frw}. In Sec.~\ref{ssect:exp}, we describe our experimental methods. In Sec.~\ref{ssect:exp_fits}, we present detailed measurements of the three-body loss rate coefficient $L_3$, which are then compared and analyzed with the UZR model. In Sec.~\ref{ssect:FiniteTCorr}, we present an empirical description of a small temperature-dependent shift of the resonance position as determined within the UZR approach. 

\subsection{Experimental procedures}\label{ssect:exp}
Our experimental setup and the procedures for preparing an ultracold cesium sample are similar to the ones reported in Refs.~\cite{Berninger2011uot,Berninger2013frw}. Here we use a single laser beam instead of two crossed laser beams to form an optical dipole trap, which contains the atoms in the lowest hyperfine and Zeeman sublevel $|F=3,m_F=3\rangle$. The near-infrared trapping beam with a waist of 40 $\mu$m is provided by a fiber laser at a wavelength of 1064 nm. Along the axial direction the trapping potential is mainly provided by the curvature of the magnetic field. A magnetic levitation gradient of $\sim 31$~G/cm is applied to compensate the gravitational force. To prepare atoms at various temperatures, we vary the trap depth at which the evaporation is stopped. An increase in trap depth of typically 50\% is applied adiabatically at the end of the evaporation to avoid unwanted evaporative losses during the hold time. By varying the final power of the trapping light between 1.2 and 260 mW, we can set the temperature in a range between about 30~nK and 3~$\mu$K. For the shallowest trap, the trap frequencies are $2\pi\times$[19.0(2), 20.8(7), 1.46(1)] Hz, where the last one is the axial frequency. For the deepest trap, the trap frequencies are $2\pi\times$[296(1), 359(3), 2.21(2)] Hz. Correspondingly, the geometric mean frequency $\bar{\omega}/2\pi$ varies between 8.3(1)~Hz and 61.7(3)~Hz. The typical atom numbers after the preparation procedure are about $5\times10^4$ for our lowest $T$ and $1.5\times10^6$ for our highest $T$.

The three-body recombination rate coefficient $L_3$ is obtained from the decay curves of the atoms hold in the dipole trap. The maximum hold time is chosen to provide a typical loss of 30\%, and thus varies between 1 and 7 s. Because of anti-evaporation~\cite{Weber2003tbr}, the temperature $T$ of the gas increases by about 10\% during the decay process. The corresponding time evolution of $T$ needs to be taken into account to extract accurate values for $L_3$. We perform time-of-flight absorption imaging at the end of the hold period to obtain the remaining atom number $N$ and the temperature $T$ at each $t$. 
For each setting of the magnetic field, both $N$ and $T$ are recorded as functions of the variable hold time $t$. 

To obtain the values for the loss-rate coefficient $L_3$, we model the atomic number evolution by the differential equation $\dot{N}/N=-3^{-3/2}L_3{(N/V)}^2$, where $V=(2\pi k_B T/m\bar{\omega}^2)^{3/2}$ is the time-dependent effective volume. The differential equation is solved with the same method as used in Ref.~\cite{Huang2014oot}. We numerically integrate the equation and fit it to the observed atom number evolution while leaving $L_3$ and the initial atom number $N_0$ as free parameters. Since our model neglects the effect of the small temperature increase on $L_3$, the results represent temperature-averaged values. We treat them as the $L_3$ values at the mean temperature. For the roughly 10\% temperature increase the resulting errors are small and we consider them as part of the uncertainties of $T$.

\subsection{Experimental results and fit analysis}\label{ssect:exp_fits}

\begin{figure}
\includegraphics[width=8.5cm]{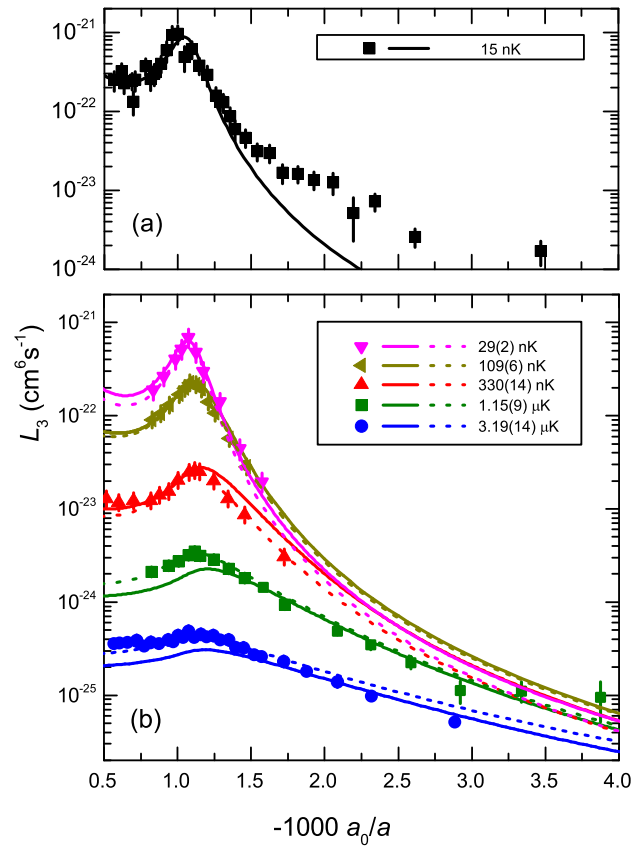}
\caption{(Color online) Three-body recombination rate coefficient $L_3$ as a function of the inverse scattering length $1/a$ at finite temperatures, measured for the ground-state Efimov resonance. Panel (a) shows the measurements at 15~nK (filled black squares) from Ref.~\cite{Berninger2011uot} and the updated fit (solid black line) from Ref.~\cite{Huang2014oot}. In panel (b), five new sets of measurements taken at different temperatures in a range between 29~nK and 3.2~$\mu$K are presented. The solid lines show the corresponding predictions from the UZR model using the resonance position and width that results from the fit to the 15~nK data in (a). The dashed lines result from individual fits based on the UZR model. The error bars show the $1\sigma$ statistic fit uncertainties.} \label{fig:logl3}
\end{figure}

An overview of all our experimental results on the temperature dependence of the ground-state Efimov resonance in the high-field region is presented in Fig.~\ref{fig:logl3}. The measured values of $L_3$ are shown as a function of the inverse scattering length $1/a$ for temperatures between 15~nK and 3.2~$\mu$K. While panel (a) shows the results at our lowest temperature from Ref.~\cite{Berninger2011uot}, panel (b) shows five new sets of measurements at higher temperatures. Our measurements clearly show that, with increasing temperature, the amplitude of the resonance decreases and the loss maximum shifts towards smaller values of $|a|$.

We now compare our experimental results with the temperature dependence according to the UZR model. The model has two free parameters, the zero-temperature position $a_-$ of the Efimov resonance and the dimensionless quantity $\eta_*$, which characterizes its width. For fitting the predictions of the model to the experimental data, we follow the strategies of our previous work~\cite{Berninger2011uot,Huang2014oot}, introducing an additional amplitude scaling factor $\lambda$ to account for systematic errors in the determination of the atomic number density.  To account for the uncertainties in our measurements of the temperature, i.e.\ deviations of the relevant temperature $T$ from the observed temperature $T_{\rm obs}$, we can alternatively use $T$ as a free fit parameter \cite{Huang2014oot}. In this case, the amplitude scaling factor is not a free parameter any more, but it is determined as $\lambda = (T_{\rm obs}/T)^3$. %Both $\lambda$ and $T$ together cannot be used as free parameters since their strong correlation renders the fits unstable. 
A comparison between the results from the two fit methods provides information on model-dependent errors. 

We follow two different strategies to compare the UZR model to our experimental data. In the first case, we fit the data set at our lowest temperature (15~nK), which is a near-ideal representation of the $T=0$ limit, to extract the resonance position or 3BP $a_-$ (for simplicity of our notation $a_- \equiv a_-^{(0)}$) and the two other parameters $\eta_*$ and $\lambda$. 
In this fit we only take into account the experimental points for $|a| > 600 a_0$ to avoid the influence of a four-body resonance~\cite{Ferlaino2009efu}, the effect of which is clearly visible in Fig.~\ref{fig:logl3}(a).
With these parameters, we then apply the UZR model to predict the corresponding $L_3$ curves for the five higher temperatures under the assumption that the 3BP is unchanged. These predictions, which are represented by the solid lines in Fig.~\ref{fig:logl3}(b), show a reasonable agreement with the experimental data. They reproduce the observed decrease of the resonant value of $L_3$ over more than three decades and they show a similar temperature shift of the maximum. A closer inspection, however, reveals significant deviations, in particular in the position of the loss maximum.

\begin{figure}
\includegraphics[width=8.5cm]{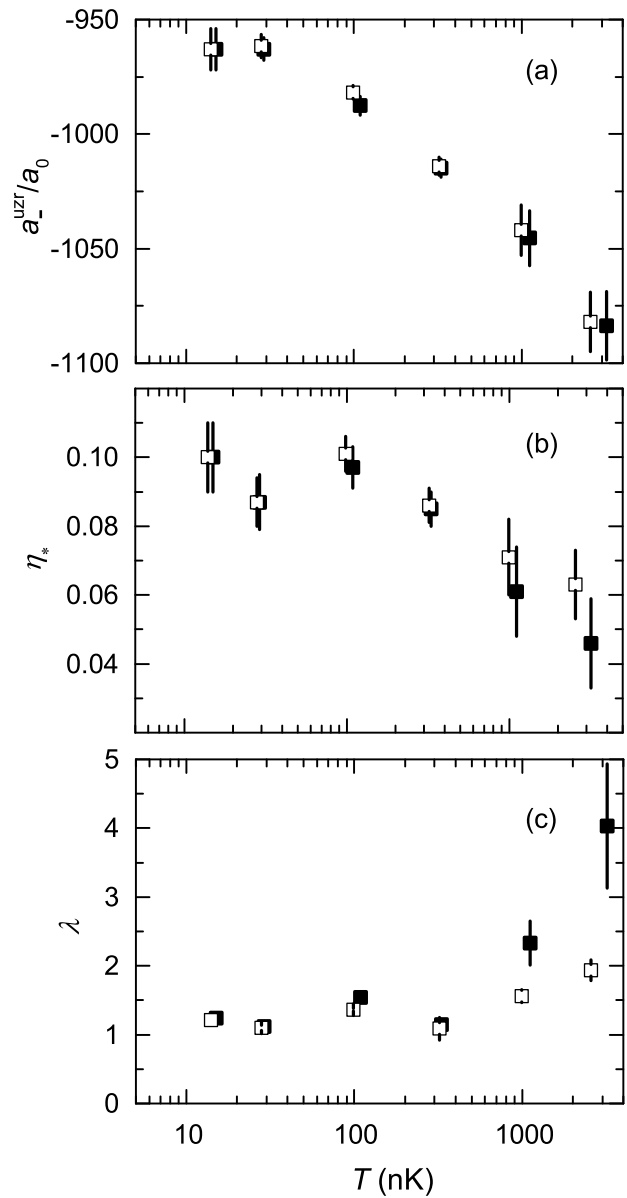}
\caption{Fit results for the UZR model applied to the experimental data of Fig.~\ref{fig:logl3}. Panels (a), (b), and (c) show $a_-^{\rm uzr}$, $\eta_*$, and $\lambda$ for the six different experimentally realized temperatures. The filled (open) squares refer to the fitting method with $T$ being fixed (free); see text. The error bars represent the $1\sigma$ statistic fit uncertainties. 
} \label{fig:params}
\end{figure}

Our second strategy is to fit all curves independently and to extract the corresponding sets of three parameters for all different temperatures separately. 
%\textcolor{blue}{Any significant temperature dependence of the fit parameters will reveal physics beyond the UZR model under the assumption of a constant 3BP.} 
%Figure~\ref{fig:params} shows the results of the independent fits as function of the temperature; here the filled (open) symbols
%correspond to the fixed-$T$ (free-$T$) fits. 
For the resonance position extracted in this way we use the notation $a_-^{\rm uzr}$ to emphasize the difference between this fit parameter of the UZR model and the true zero-temperature resonance position $a_-$. Indeed, our results in Fig.~\ref{fig:params}(a) show a clear systematic change in $a_-^{\rm uzr}$ emerging with increasing temperature. The values obtained for $\eta_*$ in (b) show relatively large uncertainties. While for lower temperatures up to about 300~nK we do not observe any significant trend, the data points at the highest two temperatures indicate a decrease of $\eta_*$ with temperature. Finally, in (c) the amplitude scaling parameter $\lambda$ also does not show any significant trend up to about 300~nK. For the highest two temperatures, however, a clear increase is observed. This appears to be rather unphysical as it is too big to be explained by our uncertainties in the atom number calibration or the temperature.

\subsection{Empirical characterization of the resonance shift}\label{ssect:FiniteTCorr}

\begin{figure}
\includegraphics[width=8.5cm]{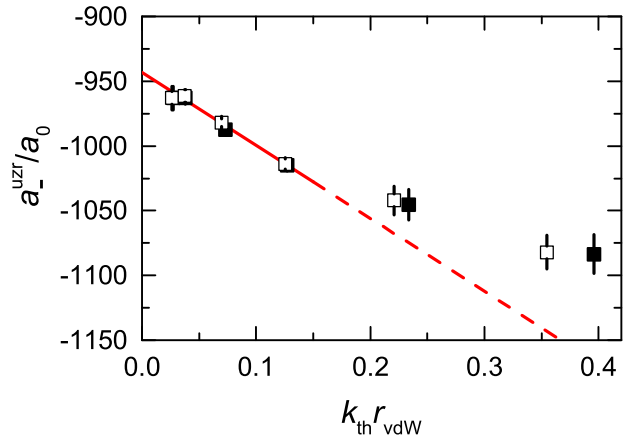}
\caption{(Color online) The resonance position parameter $a_-^{\rm uzr}$ from the UZR model fit as a function of the dimensionless quantity $k_{\rm th} r_{\rm vdW}$. The experimental data points are identical to the black squares in the panel (a) of Fig.~\ref{fig:params}. For the four lowest temperatures the data points are fitted by a straight line (solid red). The dashed line is an extrapolation to higher temperatures. The error bars show the $1\sigma$ statistic errors from fits.
} \label{fig:aminus_trend_high}
\end{figure}
%0.598(23) exp.; 0.548(55) num.

The observed temperature dependence of the fit parameter $a_-^{\rm uzr}$  can be discussed within the UZR approach as a dependence of the 3BP on the collision energy. From a different perspective, the effect may be interpreted as a consequence of the finite interaction range in real atomic systems. These two interpretations are naturally connected, since in our system the 3BP is essentially determined by the finite range of the van der Waals potential \cite{Berninger2011uot, Wang2012oot, Schmidt2012epb, Sorensen2012epa, Naidon2014moa,
Wang2014uvd}.

To characterize the effect we follow the length scale arguments outlined in  Sec.~\ref{sect:role-of-T} and introduce the dimensionless parameter $k_{\rm th} r_{\rm vdW}$. For the temperatures investigated, this quantity varies between 0.02 (at 15~nK) and 0.36 (at 3.2 $\mu$K).
We now quantify the resonance shift in an empirical way using the linear expansion
\begin{equation}\label{eq:aminus_trend}
a_-^{\rm uzr}/a_- = 1 + c\times k_{\rm th} r_{\rm vdW}.
\end{equation}

Figure~\ref{fig:aminus_trend_high} shows our results for $a_-^{\rm uzr}$ plotted as a function of $k_{\rm th} r_{\rm vdW}$. For the four lowest temperatures with $k_{\rm th} r_{\rm vdW} \lesssim 0.15$ ($T \lesssim$ 330~nK) the data points are fully consistent with a linear behavior. Only for the largest two temperatures, we observe significant deviations from the linear behavior. For these two points also the two other fit parameters $\eta_*$ and $\lambda$ show substantial deviations from the behavior in the low-temperature limit, so that there is good reason to restrict our further analysis to the four lowest temperatures. 

By fitting a straight line according to Eq.~\eqref{eq:aminus_trend} to the data points for $k_{\rm th} r_{\rm vdW} < 0.15$ (solid line in Fig.~\ref{fig:aminus_trend_high}), we extract the coefficient $c = 0.60(3)$ and the zero-temperature resonance position $a_- = -943(2) a_0$. This new value for $a_-$, obtained as an extrapolation to $T = 0$, slightly deviates from the value obtained previously~\cite{Huang2014oot} from analyzing only the set of measurements at 15~nK.

\section{Discussion}\label{sect:discussion}

The question remains to what extent we can consider the resonance shift according to Eq.~(\ref{eq:aminus_trend}) as universal. On the experimental side, this can in principle be tested by comparing it with observations on different Feshbach resonances in the same system or with other systems. On the theoretical side, the UZR model can be compared with other finite-temperature approaches that do not rely on the zero-range approximation.  Here we analyze the additional pieces of information that are available on resonances in Cs and $^6$Li and discuss the consequences of  our work in view of previous and future experiments on Efimov resonances.

\subsection{Ground-state Efimov resonance of Cs in the low-field region}
\label{ssect:Cslow}

\begin{figure}
\includegraphics[width=8.5cm]{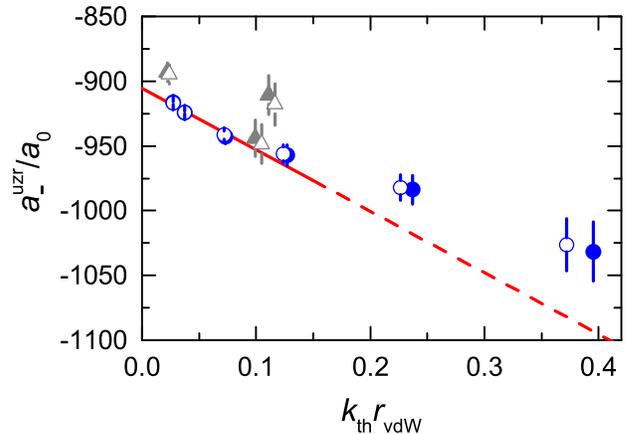}
\caption{(Color online) Analysis of the ground-state Efimov resonance in Cs at low magnetic field. The parameter $a_-^{\rm uzr}$ resulting from the UZR fit is shown as a function of the dimensionless quantity $k_{\rm th} r_{\rm vdW}$. The triangles refer to the experimental results of Ref.~\cite{Kraemer2006efe}. The circles represent the data from the numerical model of Ref.~\cite{Wang2014uvd}, and the straight line corresponds to a linear fit for $k_{\rm th} r_{\rm vdW} < 0.15$. The filled (open) symbols refer to fits with fixed $T$ (free $T$) and the error bars show the $1\sigma$ fit uncertainties.
Note the striking similarity with Fig.~\ref{fig:aminus_trend_high}.
} \label{fig:aminus_trend_low}
\end{figure}

In our early experiments on Efimov physics \cite{Kraemer2006efe}, we investigated Cs in the region of low magnetic fields, where a Feshbach resonance with very similar character as in the high-field case is available. This resonance is also strongly entrance-channel dominated, but less extremely than the high-field resonance. The main set of measurements in Ref.~\cite{Kraemer2006efe} was taken at a temperature of 10\,nK, and two further sets were recorded at 200\,nK and 250\,nK. For the present purpose we have analyzed the original $L_3$ data in the same way as described above. The results for the resonance position parameter $a_-^{\rm uzr}$ are shown by the triangles in Fig.~\ref{fig:aminus_trend_low}. Only a few data points are available with relatively large uncertainties, but they nevertheless show a clear temperature shift. We can extract a corresponding coefficient $c = 0.48(15)$, consistent with our findings for the high-field resonance. However, the few data points do not permit a test of the linearity with respect to $k_{\rm th}$.

%We find the observed temperature shift to be consistent with our findings on the high-field resonance, but because of the few data points and the large uncertainties, we cannot test the linearity with respect to $k_{\rm th}$ and a coefficient $c = 0.48(15)$ can only be extracted with a very large uncertainty. %0.48(15)

Wang and Julienne \cite{Wang2014uvd} have analyzed the situation at the low-field Feshbach resonance with a theoretical model that takes into account the van der Waals interaction and the parameters of the Feshbach resonance. The predictions of the model were found in excellent agreement with the experimental results of Ref.~\cite{Kraemer2006efe}. We have analyzed six sets of theoretical predictions produced with this model \cite{Wangpriv} for the $L_3$ coefficient in the same way as we did for the experiments. The results for  $a_-^{\rm uzr}$  are shown by the round symbols in Fig.~\ref{fig:aminus_trend_low}. The comparison of these theoretical results with the experimental results shown in Fig.~\ref{fig:aminus_trend_high} reveals a striking similarity. We find an essentially linear behavior in the range $k_{\rm th} r_{\rm vdW} < 0.15$ and by fitting a straight line (solid line) we obtain a coefficient $c = 0.55(6)$, which is fully consistent with the experimental result for the Efimov resonance in the high-field region.

This comparison suggests that the Efimov resonance in the low-field region behaves in essentially the same way as in the high-field region. The resonance shift seems to be universal, at least for different Feshbach resonances of the same character in the same atomic system.

\subsection{Excited-state Efimov resonance in $^6$Li}
\label{ssect:Li6}

Another interesting case for which temperature-dependent experimental data are available is the excited-state Efimov resonance in $^6$Li. The observation \cite{Williams2009efa} was made in a three-component spin mixture in a scenario of three overlapping Feshbach resonances, all of them with strongly entrance-channel dominated character. The results were reanalyzed in Ref.~\cite{Huang2014tbp} based on the UZR theory. For the very large length scale of an excited Efimov state, finite-range corrections can be expected to be very small. Indeed, the measurements at two different temperatures ($T = 30$\,nK corresponding to $k_{\rm th} r_{\rm vdW}  = 0.0025$ and  $T = 180$\,nK corresponding to $k_{\rm th} r_{\rm vdW}  = 0.0062$) do not reveal any significant difference.

 A straightforward ansatz to generalize Eq.~(\ref{eq:aminus_trend}) to higher-order Efimov resonances reads
\begin{equation}
a_-^{(n); \,\, {\rm uzr}}/a^{(n)}_- = 1 + c^{(n)}\times k_{\rm th} r_{\rm vdW}  \, ,
\label{eq:generalized}
\end{equation}
where $a_-^{(n); \,\, {\rm uzr}}$ denotes the resonance position obtained by the UZR fit for finite temperatures. 
Our analysis of the excited-state resonance in $^6$Li yields a coefficient $c^{(1)} = 1.6(5.8)$, which within a large uncertainty is consistent with zero.
The most simple assumption would be a constant coefficient $c^{(n)} = c$, independent of the resonance order. Within this assumption, the $^6$Li result would be consistent with $c = 0.60(3)$ as we have obtained for the Cs ground-state resonance case. In general $c^{(n)}$ can be a function of the dimensionless parameter $r_{\rm vdW}/a^{(n)}$, but because of the very large uncertainty our analysis does not provide any further information on that.

The results on the $^6$Li excited-state resonance are nevertheless very instructive as they provide a test of an alternative explanation for the observed deviations. Let us assume that there is a systematic problem with the UZR theory and the temperature-dependent shift is unrelated to the finite interaction range. In this case $r_{\rm vdW}$ would not be a relevant quantity and the problem would perfectly follow the discrete scale invariance of the Efimov problem. Then the only way to express the relative shift would be 
\begin{equation}
a_-^{(n); \,\, {\rm uzr}}/a^{(n)}_- = 1 + \tilde{c}\times k_{\rm th} |a^{(n)}|  \, .
\label{eq:nonsense}
\end{equation}
Here the relative shift of ground-state and excited-state Efimov resonances would be the same if $k_{\rm th} |a^{(n)}|$ is kept constant. Following this ansatz to analyze the data, we obtain $\tilde{c} = 0.063(3)$ for the Cs ground-state Efimov resonance and $\tilde{c} = 0.010(35)$ for the $^6$Li excited-state resonance.  These two results are inconsistent with 90\% confidence, which supports our hypothesis of a finite-range effect instead of a systematic problem in the UZR theory.

\subsection{Efimov period in Cs revisited}

A main result of Ref.~\cite{Huang2014oot} is the determination of the Efimov period as the ratio $a_-^{(1)}/a_-^{(0)} = 21.0(1.3)$, where we applied the UZR theory to both the ground-state and the excited-state Efimov resonance. The present result suggests small corrections to the positions of both resonances. From the zero-temperature extrapolation of Sec.~\ref{ssect:FiniteTCorr} we obtain the updated value $a_-^{(0)} = -943(2)\,a_0$, and Eq.~(\ref{eq:generalized}) with the assumption $c^{(1)} = c = 0.6$ yields the slightly corrected value  $a_-^{(1)}= -19930(1200)\,a_0$. In this case, the updated result for the Efimov period would be $a_-^{(1)}/a_-^{(0)} = 21.1(1.3)$. If we take the updated value for $a_-^{(0)}$, but assume there is no correction to  $a_-^{(1)}$, we obtain an Efimov period of  $21.4(1.3)$. The differences to the previous result are well within the error bar, so that the conclusions of  Ref.~\cite{Huang2014oot} remain unchanged.

\subsection{Open questions}

The considerations in Secs.~\ref{ssect:Cslow} and \ref{ssect:Li6} provide some support for a universal character of the temperature-dependent shift of the resonance position, when the UZR theory is applied to real atomic systems with a small, but finite interaction range. The generalization from the ground-state Efimov resonance to excited-state resonances according to Eq.~(\ref{eq:generalized}) raises the question on the connection between the coefficients $c^{(n)}$ for Efimov resonances of different order. Another open problem is the situation of closed-channel dominated Feshbach resonances \cite{Roy2013tot}, where the two-channel nature implies an additional length scale \cite{Petrov2004tbp} larger than the van der Waals length.  Also the role of finite-range effects related to Efimov states near Feshbach resonances of intermediate character \cite{Zaccanti2009ooa, Dyke2013frc, Gross2009oou} needs further investigations. Finally we note that in the high-temperature regime where $k_{\rm th} r_{\rm vdW}$ is no longer a small quantity, higher partial waves may substantially contribute to three-body recombination \cite{Werner2006uqt, Dincao2009tsr} and lead to non-universal behavior.

\section{conclusion}

We have investigated the temperature dependence of three-body recombination near a ground-state triatomic Efimov resonance in cesium.
Our measurements of the recombination rate coefficient extend from conditions near the zero-temperature limit to a strongly temperature-dominated regime, thus characterizing the temperature-induced resonance shift in a wide range. The two-body interactions are controlled via a broad Feshbach resonance, the character of which is strongly entrance-channel dominated.

To determine the precise zero-temperature Efimov resonance position from the finite-temperature experimental data, we have employed the universal zero-range finite-temperature theory of Refs.~\cite{Rem2013lot, Petrov2015tbr}, following our earlier investigations in Refs.~\cite{Huang2014oot, Huang2014tbp}. The present results reveal a small shift, which increases linearly with the thermal wavenumber, i.e.\ proportionally to the square root of the temperature. We attribute this effect to the finite range of the two-body interaction, which in our case is determined by the van der Waals attraction. 

A comparison with other available experimental results and a theoretical approach that explicitly takes into account the van der Waals interaction \cite{Wang2014uvd} suggests a universal character of the shift, at least for entrance-channel dominated Feshbach resonances. More work is required to understand the physics of the shift, its effect on excited-state Efimov resonances, and its implications for the precise determination of Efimov resonance positions in other systems.

\section*{Acknowledgments}

We thank Dmitry Petrov for stimulating discussions, for critical remarks on the manuscript, and for providing the source code for the UZR model. We are indebted to Yujun Wang and Paul Julienne for providing temperature-dependent predictions based on their theoretical model. We furthermore thank Martin Berninger, Alessandro Zenesini and Jesper Levinsen for fruitful discussions. We acknowledge support by the Austrian Science Fund FWF within project P23106.

%\bibliography{ultracold,HighT}

\end{document}